\documentclass[runningheads]{cl2emult}

\usepackage{makeidx}  
\usepackage{graphicx} 
\usepackage{subeqnar} 
\usepackage{multicol} 
\usepackage{cropmark} 
\usepackage{math}     
\makeindex            

\begin{document}
\def\eqn#1{Eq.$\,$#1}
\def\mb#1{\setbox0=\hbox{$#1$}\kern-.025em\copy0\kern-\wd0
\kern-0.05em\copy0\kern-\wd0\kern-.025em\raise.0233em\box0}

\title*{Statistical Mechanics of Violent Relaxation in Stellar Systems}
\toctitle{Statistical Mechanics of Violent Relaxation 
\protect\newline in Stellar Systems }
%
%
\titlerunning{Violent Relaxation in Stellar Systems}
%
\author{Pierre-Henri Chavanis}
\authorrunning{Pierre-Henri Chavanis}
%
%
\institute{Laboratoire de Physique Quantique,\\
Universit\'e Paul Sabatier,\\
118 route de Narbonne,\\
31062 Toulouse, France}

\maketitle              

\begin{abstract}

We discuss the statistical mechanics of violent relaxation in stellar
systems following the pioneering work of Lynden-Bell (1967). The
solutions of the gravitational Vlasov-Poisson system develop finer and
finer filaments so that a statistical description is appropriate to
smooth out the small-scales and describe the ``coarse-grained''
dynamics. In a coarse-grained sense, the system is expected to reach
an equilibrium state of a Fermi-Dirac type within a few dynamical
times. We describe in detail the equilibrium phase diagram and the
nature of phase transitions which occur in self-gravitating
systems. Then, we introduce a small-scale parametrization of the
Vlasov equation and propose a set of relaxation equations for the
coarse-grained dynamics. These relaxation equations, of a generalized
Fokker-Planck type, are derived from a Maximum Entropy Production
Principle (MEPP). We make a link with the quasilinear theory of the
Vlasov-Poisson system and derive a truncated model appropriate to
collisionless systems subject to tidal forces.  With the
aid of this kinetic theory, we qualitatively discuss the concept of
``incomplete relaxation'' and the limitations of Lynden-Bell's theory.

\end{abstract}

\section{Introduction}
\label{intro}

It has long been realized that galaxies, and self-gravitating systems
in general, follow a kind of organization despite the diversity of
their initial conditions and their environement. This organization is
illustrated by morphological classification schemes such as the Hubble
sequence and by simple rules which govern the structure of individual
self-gravitating systems. For example, elliptical galaxies display a
quasi-universal luminosity profile described by de Vaucouleurs'
$R^{1/4}$ law and most of globular clusters are well fitted by the
Michie-King model \cite{bt}. The question that naturally emerges is,
what determines the particular configuration to which a
self-gravitating system settles. It is possible that their present
configuration crucially depends on the conditions that prevail at
their birth and on the details of their evolution. However, in view of
their apparent regularity, it is tempting to investigate whether their
organization can be favoured by some fundamental physical principles
like those of thermodynamics and statistical physics. We ask therefore
if the actual states of self-gravitating systems are not simply more
probable than any other possible configuration, i.e. if they cannot be
considered as maximum entropy states.  This thermodynamical approach
may be particularly relevant for globular clusters and elliptical
galaxies which are described by a distribution function that is almost
isothermal
\cite{bt}. In the case of globular clusters, the relaxation proceeds
via two-body encounters and this collisional evolution is
well-described by kinetic equations of a Fokker-Planck-Landau type for
which a H-theorem 
\footnote{The H-theorem, proved by Boltzmann  
for an ideal gas, states that entropy is a monotonically increasing
function of time. Boltzmann's kinetic theory of gases therefore
provides a direct justification of the second principle of
thermodynamics. The generalization of this theorem for
self-gravitating systems rests on simplifying assumptions which are
difficult to rigorously justify
\cite{Kandrup}.} is available. By contrast, for elliptical galaxies,
two-body encounters are completely negligible (the corresponding
relaxation time $t_{{\it coll}}$ exceeds the age of the universe by
many orders of magnitude) and the galaxy dynamics is described by the
{\it Vlasov equation}, i.e. collisionless Boltzmann equation \cite{bt}. Since
the Vlasov equation rigorously conserves entropy, a relaxation towards
an isothermal distribution looks at first sight relatively
surprising. Yet, the inner regions of elliptical galaxies appear to be
isothermal and this fact stemed as a mystery for a long time.

In a seminal paper, Lynden-Bell \cite{lb} argued that the violently
changing gravitational field of a newly formed galaxy leads to a
redistribution of energies between stars and provides a mechanism
analogous to a relaxation in a gas. The importance of this form of
relaxation had previously been stressed by a number of authors
including H\'enon and King but Lynden-Bell showed for the first time
the relevance of a statistical description. He argued that the
Vlasov-Poisson system develops an intricate {\it mixing process} in
phase space associated with the heavily damped oscillations of a
protogalaxy initially far from mechanical equilibrium and collapsing
under its own gravity. As a result, the solutions of the Vlasov
equation are not smooth but involve intermingled filaments at smaller
and smaller scales. In this sense, there is no convergence towards
equilibrium but rather the formation of a fractal-like structure in
phase space.  However, if we introduce a macroscopic level of
description and make a local average of the distribution function over
the filaments, the resulting ``coarse-grained'' distribution function
is smooth and is expected to reach a maximum entropy state (i.e. most
mixed state) on a very short time scale of the order of the dynamical
time $t_{D}$. This process is called {\it violent relaxation} and is
acknowledged to account for the regularity of elliptical galaxies or
other collisionless self-gravitating systems.  Lynden-Bell predicted
that the equilibrium state should be described by a Fermi-Dirac
distribution function or a superposition of Fermi-Dirac
distributions. Here, degeneracy is due not to quantum mechanics but to
the Liouville theorem that prevents the smooth distribution function
from exceeding the maximum of its initial value. In the non degenerate
limit, the Fermi-Dirac distribution functions reduce to Maxwellians.
The prediction of an isothermal distribution for collisionless stellar
systems was considered as a triumph in the 1960's and laid the
foundation of a new type of statistical mechanics. Of course, the
validity of the theory is conditioned by a hypothesis of ergodicity
which may not be completely fulfilled. This is the complicated problem
of ``incomplete relaxation'' which limitates the power of prediction
of Lynden-Bell's approach. However, as we shall see, these difficulties
should not throw doubt on the importance of this statistical
description. A similar relaxation process is at work in
two-dimensional turbulence (described by the 2D Euler equation) and
can explain the organization and maintenance of coherent vortices, such
as the Great Red Spot of Jupiter, which are common features of
large-scale geophysical or astrophysical flows
\cite{rs1,miller,csfluide,rr,shallow,bouchet1,bouchet2}. The
mathematical relevance of this statistical description has been given
by Robert \cite{rob} introducing the concept of Young measures. The
formal analogy between two-dimensional vortices and stellar systems has
been discussed by Chavanis
\cite{cthese,cfloride,japon,csr}.

This paper is organized as follows. In section \ref{VP}, we introduce
the gravitational Vlasov-Poisson system and list its main
properties. In section \ref{LB}, we present the statistical approach
of Lynden-Bell \cite{lb} to the problem of violent relaxation. In
section \ref{FD}, we show that the Fermi-Dirac equilibrium
distribution predicted by Lynden-Bell is not entirely satisfactory
since it has an infinite mass. We must therefore invoke {\it
incomplete relaxation} and introduce truncated models. In the
artificial situation in which the system is enclosed within a
spherical box, we can calculate the Fermi-Dirac spheres explicitly and
prove the existence of a global entropy maximum for each value of
energy. For low energies, this equilibrium state has a degenerate core
surrounded by a dilute atmosphere, as calculated by Chavanis \&
Sommeria \cite{cs}. More generally, we determine the complete
equilibrium phase diagram and discuss the nature of phase transitions
in self-gravitating systems. In section \ref{MEPP}, we
describe the coarse-grained relaxation of collisionless stellar
systems towards statistical equilibrium in terms of a generalized
Fokker-Planck equation.  This relaxation equation is derived from a
phenomenological Maximum Entropy Production Principle (MEPP) and
involves a diffusion in velocity space compensated by a nonlinear
friction. In section \ref{QT}, we present a quasilinear theory of the
Vlasov-Poisson system and show that it leads to a kinetic equation of
a Landau type.  When the system is close to equilibrium (so that a
thermal bath approximation can be implemented) this equation reduces
to the Fokker-Planck equation of the thermodynamical approach and the
diffusion coefficient can be explicitly evaluated. This provides a
new, self-consistent, equation for the ``coarse-grained'' dynamics of
stellar systems where small scales have been smoothed-out in an
optimal way. In section
\ref{TM}, we use this kinetic model to derive the distribution
function of a tidally truncated collisionless stellar system. This
truncated model preserves the main features of Lynden-Bell's
distribution (including degeneracy) but has a finite mass, avoiding
the artifice of a spherical container. Other truncated models
attempting to take into account incomplete relaxation are discussed.

\section{The gravitational Vlasov-Poisson system}
\label{VP}

For most stellar systems, the encounters between stars are negligible
\cite{bt} and the galaxy dynamics is described by the self-consistent
Vlasov-Poisson system
\begin{equation}
\label{Vlasov}
{\partial f\over\partial t}+{\bf v}{\partial f\over\partial {\bf r}}+{\bf F}{\partial f\over\partial {\bf v}}=0,
\end{equation}
\begin{equation}
\label{Poisson}
\Delta\Phi=4\pi G\int f d^{3}{\bf v}.
\end{equation}
Here, $f({\bf r},{\bf v},t)$ denotes the distribution function
(defined such that $f d^{3}{\bf r}d^{3}{\bf v}$ gives the total mass of
stars with position ${\bf r}$ and velocity ${\bf v}$ at time $t$),
${\bf F}({\bf r},t)=-\nabla\Phi$ is the gravitational force (by unit
of mass) experienced by a star and $\Phi({\bf r},t)$ is the
gravitational potential related to the star density $\rho({\bf
r},t)=\int f d^{3}{\bf v}$ by the Newton-Poisson equation
(\ref{Poisson}). The Vlasov equation (\ref{Vlasov}) simply states
that, in the absence of encounters, the distribution function $f$ is
conserved by the flow in phase space. This can be written $D f/Dt=0$
where $D/Dt={\partial /\partial t}+{\bf U}_{6}\nabla_{6}$ is the
material derivative and ${\bf U}_{6}=({\bf v},{\bf F})$ is a
generalized velocity field in the $6$-dimensional phase space $({\bf
r},{\bf v})$ [by definition, $\nabla_{6}=(\partial/\partial {\bf
r},\partial/\partial {\bf v})$ is the generalized nabla
operator]. Since the flow is incompressible, i.e. $\nabla_{6}{\bf
U}_{6}=0$, the hypervolume of a ``fluid'' particle is
conserved. Since, in addition, a fluid particle conserves the
distribution function, this implies that the total mass (or
hypervolume) of all phase elements with phase density between $f$ and
$f+\delta f$ is conserved. This is equivalent to the conservation of
the Casimir integrals
\begin{equation}
\label{Casimirs}
I_{h}=\int h(f) d^{3}{\bf  r} d^{3}{\bf  v},
\end{equation}
for any continuous function $h(f)$ (they include in particular the
total mass $M=\int f d^{3}{\bf r} d^{3}{\bf v}$). It is also
straightforward to check that the Vlasov-Poisson system conserves the
total energy (kinetic $+$ potential)
\begin{equation}
\label{Energy}
E=\int{1\over 2}f v^{2} d^{3}{\bf  r}
 d^{3}{\bf  v}+{1\over 2}\int f\Phi d^{3}{\bf  r} d^{3}{\bf  v}=K+W,
\end{equation}
the angular momentum 
\begin{equation}
\label{Angular}
{\bf  L}=\int f ({\bf  r}\wedge {\bf  v}) d^{3}{\bf  r} d^{3}{\bf  v},
\end{equation}
and the impulse 
\begin{equation}
\label{Linear}
{\bf  P}=\int f {\bf  v} d^{3}{\bf  r} d^{3}{\bf  v}.
\end{equation}
In the following, we shall work in the barycentric frame of reference
and assume that the system is non rotating so that the conservation of
${\bf L}$ and ${\bf P}$ can be ignored (see Refs. \cite{lb,csr} for a
generalization).

\section{Lynden-Bell's approach of violent relaxation}
\label{LB}

When the initial condition is far from equilibrium, the Vlasov-Poisson
system develops a complicated mixing process in phase space and
generates intermingled filaments due to stirring effects. Therefore, a
{\it deterministic} description of the flow in phase space requires a
rapidly increasing amount of information as time goes on. For that
reason, it is appropriate to undertake a {\it probabilistic}
description in order to smooth out the small-scales (fine-grained) and
concentrate on the locally averaged (coarse-grained) quantities. This
statistical analysis has been considered a long time ago by
Lynden-Bell \cite{lb}. The statistical mechanics of continuous systems
is not as firmly established as in the usual case of N-body systems
but Robert \cite{rob} has recently developed a mathematical
justification of this procedure in terms of Young measures. We shall
describe below the argumentation of Lynden-Bell, which is more
intuitive in a first approach.

Starting from some arbitrary initial condition, the distribution
function is stirred in phase space but conserves its values $\eta$
(levels of phase density) and the corresponding hypervolumes
$\gamma(\eta)d\eta$ as a property of the Vlasov equation. Let us
introduce the probability density $\rho({\bf r,v},\eta)$ of finding
the level of phase density $\eta$ in a small neighborhood of the
position ${\bf r,v}$ in phase space. This probability density can be
viewed as the local area proportion occupied by the phase level $\eta$
and it must satisfy at each point the normalization condition
\begin{equation}
\label{E1}
\int\rho({\bf  r,v},\eta)d\eta=1.
\end{equation}
The locally averaged distribution function is then expressed in terms
of the probability density as
\begin{equation}
\label{E2}
\overline{f}({\bf  r,v})=\int\rho({\bf  r,v}, \eta)\eta d \eta,
\end{equation}
and  the associated (macroscopic) gravitational potential satisfies
\begin{equation}
\label{E3}
\Delta{\overline{\Phi}}=4\pi G\int \overline{f} d^{3}{\bf v}.
\end{equation}
Since the gravitational potential is expressed by space integrals of
the density, it smoothes out the fluctuations of the distribution
function, supposed at very fine scale, so $\Phi$ has negligible
fluctuations. It is then possible to express the conserved quantities
of the Vlasov equation as integrals of the macroscopic fields.  These
conserved quantities are the global probability distributions of phase
density $\gamma (\eta)$ (i.e., the total hypervolume occupied by each
level $\eta$ of phase density)
\begin{equation}
\label{E4}
\gamma(\eta)=\int\rho({\bf  r,v},\eta) d^{3}{\bf r}d^{3}{\bf v},
\end{equation}
and the total energy
\begin{equation}
\label{E5}
E=\int{1\over 2}\overline{f} v^{2} d^{3}{\bf  r}
 d^{3}{\bf  v}+{1\over 2}\int \overline{f} \ \overline{\Phi} d^{3}{\bf  r} d^{3}{\bf  v}.
\end{equation}
As discussed above the gravitational potential can be considered as
smooth, so we can express the energy in terms of the coarse-grained
functions $\overline{f}$ and $\overline{\Phi}$ neglecting the internal
energy of the fluctuations.

To determine the equilibrium distribution of the system, we need to
introduce an entropy functional like in ordinary statistical
mechanics. As is customary, Lynden-Bell defines the mixing entropy as
the logarithm of the number of microscopic configurations associated
with the same macroscopic state (characterized by the probability
density $\rho({\bf r},{\bf v},\eta)$). After introducing a counting
``\`a la Boltzmann'', he arrives at the expression \cite{lb}:
\begin{equation}
\label{E6}
S=-\int\rho({\bf  r,v},\eta)\ln\rho({\bf  r,v},\eta) d\eta d^{3}{\bf  r} d^{3}{\bf  v},
\end{equation}
where the integral extends over phase space and over all the levels of
phase elements. A mathematical justification of this entropy has been
given by Robert \cite{rob}. The most likely distribution to be reached
at equilibrium is then obtained by maximizing the mixing entropy
(\ref{E6}) subject to the constraints (\ref{E4})(\ref{E5}) and the
normalization condition (\ref{E1}). This variational problem is
treated by introducing Lagrange multipliers so that the first
variations satisfy
\begin{equation}
\label{E7}
\delta S- \beta\delta
E-\int \alpha(\eta)\delta\gamma(\eta) d\eta-\int { \zeta }({\bf
r,v}) \delta
\Biggl(\int\rho d\eta \Biggr) d^{3}{\bf  r}d^{3}{\bf  v}=0,
\end{equation}
where ${ \beta}$ is the inverse temperature and ${\alpha }(\eta)$ the
``chemical potential" of species $\eta$. The resulting optimal
probability density is a Gibbs state which has the form \cite{lb,csr}:
\begin{equation}
\label{E8}
 \rho({\bf  r,v},\eta)={e^{-\alpha(\eta)-\beta\eta ({v^{2}\over 2}+\overline{\Phi})}\over \int e^{-\alpha(\eta)-\beta\eta ({v^{2}\over 2}+\overline{\Phi})}d\eta}.
\end{equation}

The previous analysis gives a well defined procedure to compute the
statistical equilibrium states. The gravitational field is obtained by
solving the Poisson equation (\ref{Poisson}) with the distribution
function (\ref{E2}) determined by the Gibbs state (\ref{E8}). The
solution depends on the Lagrange multipliers $\beta$ and
$\alpha(\eta)$ which must be related to the conserved quantities $E$
and $\gamma(\eta)$ by equations (\ref{E5})(\ref{E4}). This precedure
determines {\it critical points} of entropy. Whether these critical
points are maxima or not is decided by the sign of the second 
order variations of entropy.

Following Lynden-Bell \cite{lb}, we consider a particular situation that
presents interesting features and for which the previous problem can
be studied in detail. Keeping only two levels $f=\eta_{0}$ and $f=0$
is convenient to simplify the discussion and is probably
representative of more general cases. Within this ``patch'' approximation, the
mixing entropy reduces to
\begin{equation}
\label{E9}
S=-\int \biggl \lbrace{\overline{f}\over\eta_{0}}\ln {\overline{f}\over\eta_{0}} +\biggl (1- {\overline{f}\over\eta_{0}}\biggr )\ln \biggl (1- {\overline{f}\over\eta_{0}}\biggr )\biggr \rbrace d^{3}{\bf r}d^{3}{\bf v},
\end{equation}
and the equilibrium coarse-grained distribution $\overline{f}=\rho({\bf r},{\bf v},\eta_{0})\eta_{0}$ takes explicitly the form
\begin{equation}
\label{E10}
\overline{f}={\eta_{0}\over 1+\lambda e^{\beta\eta_{0}\epsilon}},
\end{equation}
where $\epsilon={v^{2}\over 2}+\overline{\Phi}$ is the energy of a
star (per unit of mass) and $\lambda\equiv
e^{\alpha(\eta_{0})-\alpha(0)}>0$ an equivalent Lagrange
multiplier. Equation (\ref{E10}) is, apart from a reinterpretation of
the constants, the distribution function for the self-gravitating
Fermi-Dirac gas. Here, the exclusion principle
$\overline{f}\le\eta_{0}$ is due to the incompressibility constraint
(\ref{E1}) not to quantum mechanics. Because of the averaging
procedure, the coarse-grained distribution function can only {\it
decrease} by internal mixing, as vaccum is incorporated into the
patch, and this results in an ``effective'' exclusion
principle. Lynden-Bell's distribution therefore corresponds to a $4$-th
type of statistics since the ``fluid'' particles are distinghuishable but
subject to an exclusion principle. However, formally, this
distribution coincides with the Fermi-Dirac distribution. In the fully
degenerate case, this equilibrium has been studied extensively in
connection with white dwarf stars
\cite{chandra}. The structure of equilibrium may crucially depend  on
the degree of degeneracy (see section \ref{FD}) but Lynden-Bell \cite{lb} gives
arguments according to which stellar systems would be non degenerate
\footnote{In fact, his arguments do not apply to galactic nuclei where
his type of degeneracy may be important.}. In that limit,
$\overline{f}\ll\eta_{0}$, and equation (\ref{E10}) reduces to the
Maxwell-Boltzmann statistics
\begin{equation}
\label{E11}
\overline{f}=A e^{-\beta\eta_{0}\epsilon}.
\end{equation}
This was in fact the initial goal of Lynden-Bell in 1967: his theory
of ``violent relaxation" was able to justify a Maxwell-Boltzmann
equilibrium distribution, without recourse to collisions, on a short
time scale $\sim t_{D}\ll t_{coll}$ consistent with the age of
ellipticals. In addition, the individual mass of the stars never
appears in his theory based on the Vlasov equation. Therefore, the
equilibrium state (\ref{E11}) does not lead to a segregation by mass
in contrast with a collisional relaxation. This is in agreement with
the observed light distributions in elliptical galaxies that would
show greater colour differences if a marked segregation by mass was
established.

\section{Computation of Fermi-Dirac spheres}
\label{FD}

The first problem to be tackled is obviously the computation of
self-gravitating Fermi-Dirac spheres. Let us first consider the non
degenerate limit $\overline{f}\ll\eta_{0}$ corresponding to a dilute
system. In that case, the mixing entropy (\ref{E9}) reduces to the
ordinary Boltzmann entropy
\begin{equation}
\label{E12}
S=-\int f\ln f d^{3}{\bf r}d^{3}{\bf v},
\end{equation}
whose maximization at fixed mass and energy leads to the
Maxwell-Boltzmann statistics (\ref{E11}). Substituting this optimal
distribution in the Poisson equation (\ref{Poisson}), we find that the
gravitational potential is determined at equilibrium by the
differential equation
\begin{equation}
\label{E13}
\Delta\Phi=4\pi G A' e^{-\beta \Phi},
\end{equation} 
where the Lagrange multipliers $A'$ and $\beta$ have to be related to
the total mass and total energy of the system. The Boltzmann-Poisson
equation (\ref{E13}) has been studied extensively in the context of
isothermal gaseous spheres \cite{chandra} and in the case of
collisional stellar systems such as globular clusters \cite{lbw}. We
can check by direct substitution that the distribution
\begin{equation}
\label{E14}
\Phi_{s}(r)={1\over\beta}\ln (2\pi G\beta A r^{2}),\qquad\qquad \rho_{s}(r)={1\over 2\pi G\beta r^{2}},
\end{equation}
is an exact solution of equation (\ref{E13}) known as the {\it
singular isothermal sphere} \cite{bt}. Since $\rho\sim r^{-2}$ at
large distances, the total mass of the system $M=\int_{0}^{+\infty}
\rho 4\pi r^{2}dr$ is infinite! More generally, we can show that any
solution of the Boltzmann-Poisson equation (\ref{E13}) behaves like
the singular sphere as $r\rightarrow +\infty$ and has therefore an
infinite mass. This means that no equilibrium can exist in an
unbounded domain: the density can spread indefinitely while conserving
energy and increasing entropy \cite{bt}.

In practice, the infinite mass problem does not arise if we realize that
the relaxation of the system is necessarily {\it incomplete}
\cite{lb}. There are two major reasons for incomplete relaxation: (i)
The mean field relaxation process is dependant on the strength of the
variations in potential. As these die out, the relaxation ceases and
it is likely that the system may find a stable steady state before the
relaxation process is completed. Therefore, the relaxation is
effective only in a finite region of space (roughly the main body of
the galaxy) and during a finite period of time (while the galaxy is
dynamically unsteady). Orbits which lie partly outside the relaxing
region and have periods longer than the time for which the galaxy is
unsteady will not acquire their full quotas of stars. (ii) in practice,
the galaxy will not be isolated but will be subject to the tides of
other systems. Therefore, high energy stars will escape the system
being ultimately captured by the gravity of a nearby object. These two
independant effects have similar consequences and will produce a
modification of the distribution function at high energies. Some
truncated models can be obtained by developing a kinetic theory of
encounterless relaxation (sections \ref{MEPP}-\ref{TM}). This kinetic
approach will provide a precise framework to understand what limits
relaxation and why complete equilibrium is not reached in general.

\begin{figure}
\centering
\includegraphics[width=0.9\textwidth]{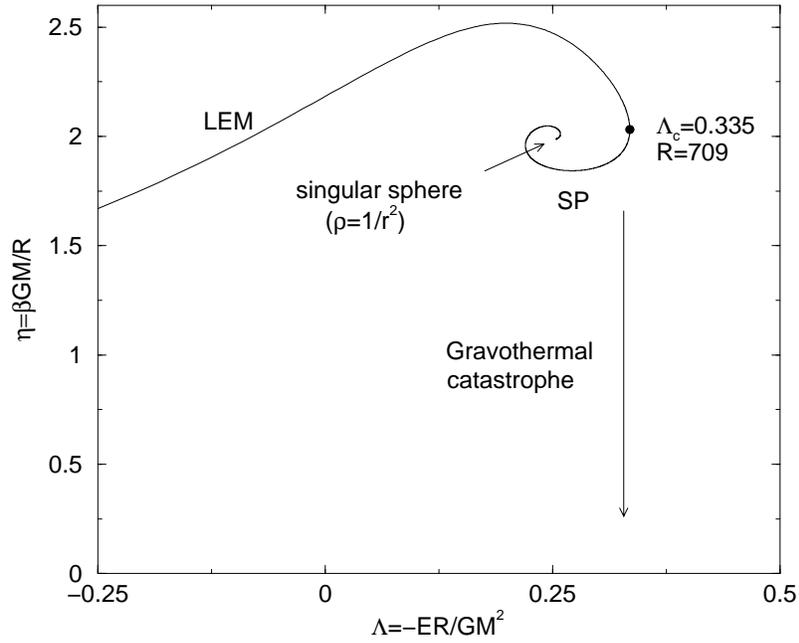}
\caption[]{Equilibrium phase diagram for classical isothermal spheres.  }
\label{figure1}
\end{figure}

However, for the present, we shall avoid the infinite mass problem by
confining artificially the system within a box of radius $R$. The
calculation of finite isothermal spheres has been carried out by
Antonov \cite{antonov}, Lynden-Bell \& Wood \cite{lbw}, Katz
\cite{katz}, Padmanabhan \cite{pad2,pad}, de Vega \& Sanchez
\cite{vega4} and Chavanis \cite{chavacano} (see Chavanis \cite{chavarelat} 
for an extension in general relativity). These studies were performed
in the framework of gaseous stars or collisional stellar systems
(e.g., globular clusters) but they extend in principle to violently
relaxed collisionless stellar systems described by the distribution
function (\ref{E11}). Therefore, we shall give a brief summary of
these classical results before considering the case of the Fermi-Dirac
distribution (\ref{E10}). The phase diagram is represented in
Fig. \ref{figure1} where the inverse temperature is plotted as a
function of minus the energy. It is possible to prove the following
results: (i) there is no global maximum for the Boltzmann
entropy. (ii) there are not even critical points for the Boltzmann
entropy if $\Lambda=-ER/GM^{2}>0.335$. (iii) local entropy maxima
(LEM) exist if $\Lambda=-ER/GM^{2}<0.335$; they have a density
contrast ${\cal R}=\rho(0)/\rho(R)<709$ (upper branch). (iv) critical
points of entropy with density contrast ${\cal R}>709$ are unstable
saddle points SP (spiraling curve). Conclusions (i) and (ii) have been
called ``gravothermal catastrophe" or ``Antonov instability''. When
$\Lambda=-ER/GM^{2}>0.335$, there is no hydrostatic equilibrium and
the system is expected to collapse and overheat. This is a natural
evolution (in a thermodynamical sense) because a self-gravitating
system can always increase entropy by taking a ``core-halo'' structure
and by making its core denser and denser (and hotter and hotter) \cite{bt}. As
discussed by Lynden-Bell \& Wood \cite{lbw}, this instability is
probably related to the negative specific heats of self-gravitating
systems: by losing heat, the core grows hotter and evolves away from
equilibrium.

The ``gravothermal catastrophe'' picture has been confirmed by
sophisticated numerical simulations of globular clusters that
introduce a precise description of heat transfers between the ``core''
and the ``halo'' using moment equations
\cite{larson}, orbit averaged Fokker-Planck equation \cite{cohn} or
fluid equations for a thermally conducting gas \cite{lbe}. In these
studies, the collapse proceeds self-similarly (with power law
behaviours) and the central density becomes infinite in a finite
time. This singularity has been known as ``core collapse'' and many
globular clusters have probably experienced core collapse. This is
certainly the most exciting theoretical aspect of the collisional
evolution of stellar systems. In practice, the formation of hard
binaries can release sufficient energy to stop the collapse and even
drive a reexpansion of the system \cite{ilb}. Then, a series of
gravothermal oscillations should follow but it takes times much larger 
than the age of globular clusters \cite{bett}. It has to be
noted that, although the central density tends to infinity, the mass
contained in the core tends to zero so, in this sense, the
gravothermal catastrophe is a rather unspectacular process \cite{bt}.
However, the gravothermal catastrophe may also be at work in dense
clusters of compact stars (neutron stars or stellar mass black holes)
embedded in evolved galactic nuclei and it can presumably initiate the
formation of supermassive black holes via the collapse of such
clusters. Indeed, when the central redshift (related to the density
contrast) exceeds a critical value $z_{c}\sim 0.5$, a dynamical
instability of general-relativistic origin sets in and the star
cluster undergoes a catastrophic collapse to a black hole on a
dynamical time \cite{rasio}. It has been suggested that this process
leads naturally to the birth of supermassive black holes of the right
size to explain quasars and AGNs \cite{st}.

Since collisionless stellar systems undergoing violent relaxation are
described by an {\it isothermal} distribution function of
the Fermi-Dirac type (\ref{E10}), it is particularly relevant to
ask whether galaxies in their birth stages can undergo a form of
gravothermal catastrophe and if so what they do about it.  The
collapse of galaxies was described qualitatively by Lynden-Bell \&
Wood \cite{lbw}: ``...The centers will then begin to separate into a
core - a sort of separate body which might even be called a
nucleus. This will cease to shrink when it becomes degenerate in
Lynden-Bell's sense. The system will then have a core-halo structure
which is an equilibrium of an isothermal Fermi-Dirac gas sphere. These
will show a variety of different nuclear concentrations depending on
the degeneracy parameter. It is evident that theory developed along
these lines has the chance of making sense of the variety of different
galactic nuclei.''

\begin{figure}
\centering
\includegraphics[width=0.9\textwidth]{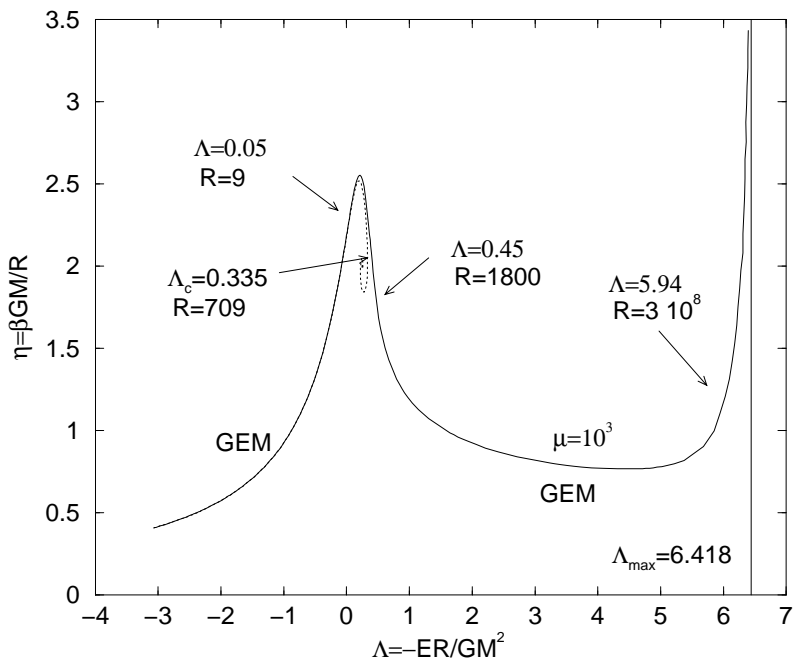}
\caption[]{Equilibrium phase diagram for Fermi-Dirac spheres ($\mu=10^{3}$).}
\label{figure2}
\end{figure}

\begin{figure}
\centering
\includegraphics[width=0.9\textwidth]{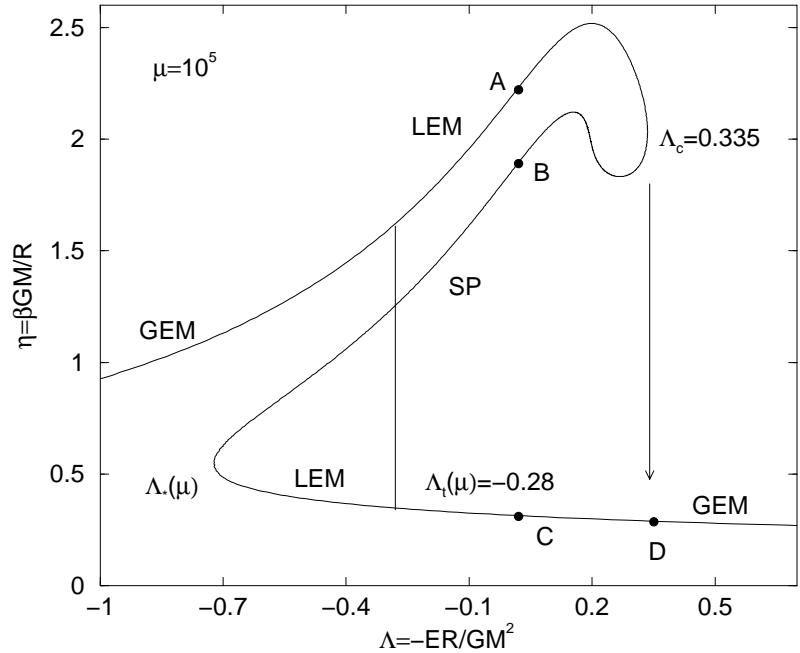}
\caption[]{Equilibrium phase diagram for Fermi-Dirac spheres ($\mu=10^{5}$).}
\label{figure3}
\end{figure}

The computation of isothermal Fermi-Dirac spheres has been performed
only recently by Chavanis \& Sommeria \cite{cs}. It can be proved that
a global entropy maximum
exists for all accessible values of energy
\cite{cs,robert}. Therefore, degeneracy has a stabilizing role and is
able to stop the ``gravothermal catastrophe''. The equilibrium diagram
is represented in Figs. \ref{figure2}-\ref{figure3} and now depends
on the degeneracy parameter $\mu=\eta_{0}/\langle f \rangle\equiv
\eta_{0}\sqrt {512\pi^{4} G^{3}MR^{3}}$, where $\eta_{0}$ is the 
maximum allowable value of the distribution function and $\langle
f\rangle$ its typical average value in the box of radius $R$. We see
that the inclusion of degeneracy has the effect of unwinding the
spiral (dashed line). When $\mu$ is small (Fig. \ref{figure2}), there
is only one critical point of entropy for each value of energy and it
is a global entropy maximum (GEM). For small values of $\Lambda$ (high
energies) the solutions almost coincide with the classical, non
degenerate, isothermal spheres. When $\Lambda$ is increased (low
energies) the solutions take a ``core-halo'' structure with a
partially degenerate core surrounded by a dilute Maxwellian
atmosphere. It is now possible to overcome the critical energy
$\Lambda_{c}=0.335$ and the critical density contrast ${\cal R}\simeq
709$ found by Antonov for a classical isothermal gas.  In that region,
the specific heat $C=dE/dT$ is negative.  As energy decreases further,
more and more mass is concentrated into the nucleus (which becomes
more and more degenerate) until a minimum accessible energy,
corresponding to $\Lambda_{max}(\mu)$, at which the nucleus contains
all the mass. In that case, the atmosphere has been ``swallowed'' and
the system has the same structure as a cold white dwarf star
\cite{chandra}.  This is a relatively singular limit since the density
drops to zero at a finite radius whereas for partially degenerate
systems, the density decays like $r^{-2}$ at large distances. When the
degeneracy parameter $\mu$ is large (Fig.
\ref{figure3}), there are now several critical points of entropy for
each single value of energy in the range $\Lambda_{*}(\mu) <\Lambda
<\Lambda_{c}$. The solutions on the upper branch (points A) are non
degenerate and have a smooth density profile; they form the
``gaseous'' phase. The solutions on the lower branch (points C) have a
``core-halo'' structure with a degenerate nucleus and a dilute
atmosphere; they form the ``condensed'' phase. According to the
theorem of Katz \cite{katz}, they are both entropy maxima while the
intermediate solutions, points B, are unstable saddle points (SP). These
points are similar to points A, except that they contain a small
embryonic nucleus (with small mass and energy) which plays the role of
a ``germ'' in the langage of phase transitions. For small values of
$\Lambda$, the non degenerate solutions (upper branch) are global
entropy maxima (GEM). They become local entropy maxima (metastable
states) at a certain $\Lambda_{t}\ge
\Lambda_{*}$ depending on the degeneracy parameter. At that
$\Lambda_{t}$, the degenerate solutions (lower branch) that were only
local entropy maxima (LEM) suddenly become global entropy maxima. We
expect therefore a phase transition to occur between the ``gaseous''
phase (upper branch) and the ``condensed'' phase (lower branch) at
$\Lambda_{t}$. For $\Lambda_{t}(\mu)<\Lambda<\Lambda_{c}$, the non
degenerate solutions (points A) are metastable but we may suspect
\cite{cs} that they are long-lived so that they {\it are}
physical. It is plausible that these metastable states will be
selected by the dynamics even if states with higher entropy
exist. This depends on a complicated notion of ``basin of
attraction'' as studied by Chavanis, Rosier and Sire \cite{crs} with
the aid of a simple model of gravitational dynamics. Therefore, the
true phase transition will occur at $\Lambda_{c}=0.335$ at which the
gaseous phase completely disappears: at that point, the gravothermal
catastrophe sets in but, for collisionless stellar systems (or for
fermions), the core ultimately ceases to shrink when it becomes
degenerate. In that case, the system falls on to a global entropy
maximum which is the true equilibrium state for these systems (point
D). This global entropy maximum has a ``core-halo'' structure with a
degenerate core and a dilute atmosphere. A simple analytical model of
these phase transitions has been proposed in Ref. \cite{pt} and provides 
a fairly good agreement with the full numerical solution. A particularity
of self-gravitating systems, which are in essence non-extensive, is
that the statistical ensembles (microcanonical and canonical) are not
interchangeable. Therefore, the description of the equilibrium diagram
is different whether the system evolves at fixed energy of fixed
temperature. A  discussion of this interesting phenomenon can be found
in the review of Padmanabhan \cite{pad} and in Chavanis \cite{pt}.

For astrophysical purposes, it is still a matter of debate to decide
whether collisionless stellar systems like elliptical galaxies are
degenerate (in the sense of Lynden-Bell) or not. Since degeneracy can
stabilize the system without changing its overall structure at large
distances, we have suggested that degeneracy could play a role in
galactic nuclei \cite{cs}. The recent simulations of Leeuwin
and Athanassoula \cite{leeuwin} and the theoretical model of Stiavelli
\cite{sti} are consistent with this idea especially if the nucleus of
elliptical galaxies contains a primordial massive black hole. Indeed,
the effect of degeneracy on the distribution of stars surrounding the
black hole can explain the cusps observed at the center of galaxies
\cite{sti}.  This form of degeneracy is also relevant for massive
neutrinos in Dark Matter models where it competes with quantum
degeneracy \cite{kull}. In fact, the Fermi-Dirac distribution of
massive neutrinos in Dark Matter models (which form a collisionless
self-gravitating system) might be justified more by the process of
``violent relaxation'' \cite{lb} than by quantum mechanics
\cite{ruffini,bilic}. As shown by Chavanis \& Sommeria \cite{cs}, 
violent relaxation can lead to the formation of a dense
degenerate nucleus with a small radius and a large mass.
This massive degenerate nucleus could be an alternative to black holes
at the center of galaxies \cite{bilic,bilic2,cs} .

\section{The Maximum Entropy Production Principle}
\label{MEPP}

Basically, a collisionless stellar system is described in a
self-consistent mean field approximation by the Vlasov-Poisson system
(\ref{Vlasov})(\ref{Poisson}). In principle, these coupled equations
determine completely the evolution of the distribution function
$f({\bf r},{\bf v},t)$. However, as discussed in section \ref{LB}, we
are not interested in practice by the finely striated structure of the
flow in phase space but only by its macroscopic, i.e. smoothed-out,
structure. Indeed, the observations and the numerical simulations are
always realized with a finite resolution. Moreover, the
``coarse-grained'' distribution function $\overline{f}$ is likely to
converge towards an equilibrium state contrary to the exact
distribution $f$ which develops smaller and smaller scales.

If we decompose the distribution function and the gravitational
potential in a mean and fluctuating part ($f=\overline{f}+\tilde f$,
$\Phi=\overline{\Phi}+\tilde\Phi$) and take the local average of the
Vlasov equation (\ref{Vlasov}), we readily obtain an equation of the
form
\begin{equation}
\label{E15}
{\partial\overline{f}\over\partial t}+{\bf v}{\partial \overline{f}\over\partial {\bf r}}+
\overline{\bf F}{\partial \overline{f}\over\partial {\bf v}}=
-{\partial {\bf J}_{f}\over\partial {\bf v}},
\end{equation}
for the ``coarse-grained'' distribution function with a diffusion
current ${\bf J}_{f}=\overline{\tilde f\tilde {\bf F}}$ related to the
correlations of the ``fine-grained'' fluctuations.

The problem in hand consists in determining a closed form for the
diffusion current ${\bf J}_{f}$. Its detailed expression
$\overline{\tilde f\tilde {\bf F}}$ depends on the subdynamics and is
therefore extremely difficult to capture from first
principles. Indeed, the ``violent relaxation'' is a very nonlinear and
very chaotic process ruling out any attempt to implement {\it
perturbative} methods if we are far from equilibrium. For that reason,
there is a strong incentive to explore {\it variational} methods which
are considerably simpler and give a more intuitive understanding of
the problem. We propose to describe the relaxation of the
coarse-grained distribution function $\overline{f}$ towards the Gibbs
state (\ref{E8}) with a Maximum Entropy Production Principle
\cite{csr}. This thermodynamical principle assumes that: ``during its
evolution, the system tends to maximize its rate of entropy production
$\dot S$ while satisfying all the constraints imposed by the
dynamics''.  There is no rigorous justification for this principle and
it is important therefore to confront the MEPP with kinetic theories
(when they are available) in order to determine its range of validity. In
any case, the MEPP can be considered as a convenient tool to build
relaxation equations which are mathematically well-behaved and which
can serve as numerical algorithms to calculate maximum entropy
states. It is also ``perfectible'' in the sense that any new
information about the dynamics of the system can be incorporated. This
principle is reminiscent of Jaynes \cite{jaynes} ideas and is a clear
extension of the well-known principle of statistical mechanics
according to which: ``at equilibrium, the system is in a maximum
entropy state consistent with all the constraints''. This is also the
most natural extension of Lynden-Bell's theory out of equilibrium.

Like for equilibrium states, the evolution of the flow in phase space
is described by a set of local probabilities $\rho({\bf r,v},\eta,t)$
and the locally averaged gravitational potential $\overline{\Phi}({\bf
r},t)$ is obtained by the integration of equation (\ref{Poisson})
where $f({\bf r,v},t)$ is replaced by $\overline{f}({\bf
r,v},t)=\int\rho({\bf r,v},\eta,t)\eta d\eta$. The phase elements are
thus transported in phase space by the corresponding averaged velocity
field $\overline{\bf U}_{6}=({\bf v},-\nabla\overline{\Phi})$ and we
suppose that, in addition, they undergo a diffusion process. This
diffusion occurs only in velocity space, due to the fluctuations of
the gravitational field ${\Phi}$. There is no diffusion in position
space since the velocity ${\bf v}$ is a pure coordinate and does not
fluctuate. As a result, the probability densities will satisfy a
convection-diffusion equation of the form
\begin{equation}
\label{E16}
{\partial \rho\over \partial t}+{\bf  v}{\partial\rho\over\partial{\bf  r}}+{\overline {\bf F}}{\partial \rho\over\partial{\bf  v}}=-{\partial {\bf  J}\over\partial{\bf  v}},
\end{equation}
where ${\bf J}({\bf r,v},\eta,t)$ is the diffusion current of the
phase element $\eta$. This equation conserves the total hypervolume
(in phase space) occupied by each phase element. Multiplying equation
(\ref{E16}) by $\eta$ and integrating over the levels of phase density
returns equation (\ref{E15}) with ${\bf J}_{f}=\int {\bf J}({\bf
r,v},\eta,t)\eta d\eta$.

To apply the MEPP, we need first to compute the rate of change of
entropy during the convection-diffusion process. Taking the time
derivative of equation (\ref{E6}), expressing $\partial_{t}\rho$ by
equation (\ref{E16}) and noting that $\rho\ln\rho$ is conserved by the
advective term, we get
\begin{equation}
\label{E17}
\dot S=-\int{\bf  J}{\partial \ln \rho\over\partial{\bf  v}} d^{3}{\bf  r} d^{3}{\bf  v} d\eta.
\end{equation}
We now determine ${\bf J}$ such that, for a given field $\rho$ at each
time $t$, ${\bf J}$ maximizes the entropy production $\dot S$ with
appropriate dynamical constraints, which are: -the conservation of the
local normalization condition (\ref{E1}) implying
\begin{equation}
\label{E18}
\int {\bf  J}({\bf  r,v},\eta,t)d \eta=0,
\end{equation}
-the conservation of energy expressed from equations (\ref{E5}) and
(\ref{E15}) as
\begin{equation}
\label{E19}
\dot E=\int {\bf  J}_{f}{\bf  v} d^{3}{\bf  r} d^{3}{\bf  v}=0,
\end{equation}
- a limitation on the eddy flux $|{\bf J}|$, characterized by a bound
$C({\bf r},{\bf v},t)$, which exists but is not specified
\begin{equation}
\label{E20}
\int{J^{2}\over 2\rho}d\eta\le C({\bf  r,v},t).
\end{equation}
This variational problem can be solved by introducing (at each time
$t$) Lagrange multipliers ${\mb\zeta}$, $\beta$, $1/D$ for the three
respective constraints. It can be shown by a convexity argument that
reaching the bound (\ref{E20}) is always favorable for increasing
$\dot S$, so that this constraint can be replaced by an equality. The
condition
\begin{equation}
\label{E21}
\delta \dot S- \beta\delta \dot E-\int {1\over D}\delta \biggl(\int {J^{2}\over 2\rho}d\eta\biggr)d^{3}{\bf  r} d^{3}{\bf  v}-\int {\mb\zeta } \delta\biggl(\int {\bf  J}d\eta\biggr)d^{3}{\bf  r} d^{3}{\bf  v}=0,
\end{equation}
yields an optimal current of the form \cite{csr}:
\begin{equation}
\label{E22}
{\bf  J}=- D({\bf  r,v},t)\biggl \lbrack{\partial\rho\over\partial{\bf  v }}+\beta(\eta-\overline{f})\rho {\bf  v}\biggr \rbrack.
\end{equation}
The Lagrange multiplier ${\mb \zeta}$ has been eliminated, using the
condition (\ref{E18}) of local normalization conservation. At
equilibrium, the diffusion currents must vanish and we can check that
this yields the Gibbs state (\ref{E8}) with $\beta$ the corresponding
inverse temperature \cite{csr}. During the evolution, this quantity
varies with time and is determined by the condition of energy
conservation (\ref{E19}). Introducing (\ref{E22}) in the constraint
(\ref{E19}), we get
\begin{equation}
\label{E23}
\beta(t)=-{\int D{\partial \overline{f}\over\partial {\bf v}}{\bf v}d^{3}{\bf r}d^{3}{\bf v}\over \int D (\overline{f^{2}}-\overline{f}^{2})v^{2} d^{3}{\bf r}d^{3}{\bf v}},
\end{equation}
where $\overline{f^{n}}=\int \rho\eta^{n}d\eta$ are the local moments of the
fine-grained distribution function. We have thus obtained a complete
set of relaxation equations which exactly conserve the distribution of
phase levels and energy.  The rate of entropy production can be put in
the form \cite{csr}:
\begin{equation}
\label{E24}
\dot S=\int {J^{2}\over D\rho}d^{3}{\bf r}d^{3}{\bf v}d\eta,
\end{equation}
so the diffusion coefficient $D$ must be positive for entropy
increase. Except for its sign, the diffusion coefficient is not
determined by the thermodynamical approach as it is related to the
unknown bound $C({\bf r},{\bf v},t)$.

These relaxation equations (\ref{E16})(\ref{E22})(\ref{E23}) can be
simplified in the case of a single non zero density level $\eta_{0}$
and provide a new non linear equation for the evolution of the
coarse-grained distribution function $\overline{f}=\rho\eta_{0}$ :
\begin{equation}
\label{E25}
{\partial \overline{f}\over \partial t}+{\bf  v}{\partial\overline{f}\over\partial{\bf  r}}+{\bf F}{\partial\overline{f}\over\partial{\bf  v}}={\partial\over\partial{\bf  v}}\biggl\lbrack D \biggl ({\partial\overline{f}\over\partial{\bf  v }}+\beta\overline{f} (\eta_{0}-\overline{f}){\bf  v}\biggr )\biggr\rbrack.
\end{equation}
In the non degenerate limit ($\overline{f}\ll \eta_{0}$), equation (\ref{E25}) takes the form of a Fokker-Planck equation
\begin{equation}
\label{E26}
{\partial \overline{f}\over \partial t}+{\bf  v}{\partial\overline{f}\over\partial{\bf  r}}+{\bf F}{\partial\overline{f}\over\partial{\bf  v}}={\partial\over\partial{\bf  v}}\biggl\lbrack D \biggl ({\partial\overline{f}\over\partial{\bf  v }}+\beta\eta_{0}\overline{f}{\bf  v}\biggr )\biggr\rbrack.
\end{equation}
This Fokker-Planck equation (sometimes called the
Kramers-Chandrasekhar equation) is well-known in the case of
collisional stellar systems (without the bar on $f$!) and is usually
derived from a Markov hypothesis and a stochastic Langevin equation
\cite{chandrabrownien}. It can also be obtained from the following
argument: because of close encounters, the stars undergo brownian
motion and diffuse in velocity space (responsible for the first term
in the right hand side). However, under the influence of this
diffusion, the kinetic energy per star will diverge as $<{v^{2}/
2}>\sim 3Dt$ and one is forced to introduce an ``ad hoc" dynamical
friction (second term in the right hand side) with a friction
coefficient $\xi=\beta D\eta_{0}$ (Einstein relation) in order to
compensate this divergence and recover the Maxwellian distribution of
velocities at equilibrium. From these two considerations (which
express the fluctuation-dissipation theorem) results the ordinary
Fokker-Planck equation (\ref{E26}). The Fokker-Planck equation is
justified here by an argument of a very wide scope that does not
directly refer to the subdynamics: it appears to maximize the rate of
entropy production at each time with appropriate
constraints. Therefore, it applies equally well to collisional or
(coarse-grained) collisionless stellar systems. In this description,
the diffusion term and the ``dynamical friction'' directly result from
the variational principle and are associated with the variations of
$S$ and $E$ respectively. Furthermore, the inverse temperature $\beta$
appears as a Lagrange multiplier associated with the conservation of
energy and the Einstein relation is automatically satisfied. In
addition, our procedure can take into account the degeneracy of
collisionless stellar systems, keeping equation (\ref{E25}) instead of
the non degenerate limit (\ref{E26}). In the degenerate case, the
friction is non linear in $\overline{f}$ so that equation (\ref{E25})
is {\it not} a Fokker-Planck equation in a strict sense. This
nonlinearity is necessary to recover the Fermi-Dirac distribution
function at equilibrium (while a linear friction drives the system
towards the Maxwell-Boltzmann distribution). This generalization is
important because degeneracy is specific to collisionless systems and
may be crucial for the existence of an equilibrium state (see section
\ref{FD}). 

In equations (\ref{E25}) (\ref{E26}), $\beta$ is not constant but
evolves with time so as to conserve the total energy. In the non
degenerate limit, and assuming that $D$ is constant, we get after a
part integration
\begin{equation}
\label{E28}
\beta(t)=-{\int {\partial \overline{f}\over\partial {\bf v}}{\bf v}d^{3}{\bf r}d^{3}{\bf v}\over \int \eta_{0}\overline{f} v^{2} d^{3}{\bf r}d^{3}{\bf v}}={3M\over 2\eta_{0}K(t)}
\end{equation}
This equation relates the formal Lagrange multiplier $\beta(t)$ to the
inverse of the {\it average} kinetic energy. This is of course satisfying on
a physical point of view.  An alternative Fokker-Planck equation
involving a {\it local} temperature $T({\bf r},t)$ instead of
$\beta(t)^{-1}$ has been proposed by Clemmow \& Dougherty \cite{cd} in
the case of collisional systems. The energy is assumed to be locally
conserved by the collisions, which is valid when the mean free path is
much smaller than the size of the system. By contrast, this hypothesis
does not seem to be justified for the violent relaxation of a
collisionless system, which is rather a global process.

Let us briefly review the main properties of our relaxation
equations. First of all, they rigorously satisfy the conservation of
energy and phase space hypervolumes like the Vlasov-Poisson system
(the conservation of impulse and angular momentum can also be
satisfied by introducing appropriate Lagrange multipliers
\cite{csr}). Moreover, they guarantee the increase of entropy at each
time $(\dot S\ge 0)$ with an optimal rate. Of course, this H-theorem
is true for the coarse-grained entropy $S=-\int\rho\ln\rho d^{3}{\bf
r}d^{3}{\bf v}d\eta$ and not for the fine-grained entropy
$S_{f.g}=-\int f\ln f d^{3}{\bf r}d^{3}{\bf v}$ which is constant, as
the integral of any function of $f$. The source of irreversibility is
due to the coarse-graining procedure that smoothes out the
small-scales and erases the microscopic details of the
evolution. Accordingly, the relaxation equations
(\ref{E16})(\ref{E22})(\ref{E23}) are likely to drive the system
towards an equilibrium state (the Gibbs state (\ref{E8})) contrary to
the Vlasov equation that develops finer and finer scales. In
mathematical terms, this means that the distribution function $f$
converges to an equilibrium state $\overline{f}$ {\it in a weak
sense}. In fact, the situation is more complicated since the Gibbs
state does not exist in an infinite domain (section \ref{FD}). We
shall see, however, in section \ref{TM} that the diffusion coefficient
is proportional to the fluctuations of the distribution function
(equation (\ref{E65})). Now, in physical situations, these
fluctuations vanish before the systems had time to relax
completely. As a result, the relaxation stops and the system remains
frozen in a subdomain of phase space. It is only in this subdomain
(corresponding to the main body of the galaxy) that the Gibbs state is
justified. This provides a physical mechanism for confining galaxies
and justifying truncated models. Alternatively, if the galaxy is not
isolated but subject to the tides of a neighboring object, a tidally
truncated model can be explicitly derived from these relaxation
equations (section \ref{TM}). It has a finite mass while preserving
the essential features of Lynden-Bell's distribution function.

The relaxation equations (\ref{E16})(\ref{E22})(\ref{E23}) can be
simplified in two particular limits when: (i) the initial condition is
approximated by a single level of phase density $\eta_{0}$ surrounded
by vaccum (ii) degeneracy, in Lynden-Bell's sense, is neglected. These
simplifications lead to the Fokker-Planck equation (\ref{E26}). If the
galaxy has sufficiently large energy, this equation will drive the
system towards a Maxwell-Boltzmann equilibrium state (\ref{E11}).
However, if the energy falls below a critical value, the Fokker-Planck
equation does not reach any equilibrium state anymore and the system
can achieve ever increasing values of entropy by developing core
collapse (Antonov instability). For collisionless stellar systems,
this ``gravothermal catastrophe'' should stop when the center of the
galaxy becomes degenerate (see section \ref{FD}). In that case, the
Fokker-Planck equation (\ref{E26}) is not valid anymore and must be
replaced by the degenerate relaxation equation (\ref{E25}). This
equation converges towards the Fermi-Dirac equilibrium state
(\ref{E10}), which exists for all values of energy.

In summary, the MEPP is able to yield relevant relaxation equations
for the coarse-grained dynamics of collisionless stellar systems
experiencing violent relaxation. This relatively elegant and simple
variational principle shows that the global structure of the
relaxation equations is determined by purely thermodynamical
arguments. All explicit reference to the subdynamics is encapsulated
in the diffusion coefficient which cannot be captured by the MEPP (it
appears as a Lagrange multiplier related to an unknown bound on the
diffusion flux). It must be therefore calculated with a kinetic 
model such as the quasilinear theory described in the next section.

\section{The quasilinear theory}
\label{QT}

The quasilinear theory of the Vlasov-Poisson system was first
considered by Kadomtsev \& Pogutse \cite{kp} for a homogeneous
Coulombian plasma and extended by Severne \& Luwel \cite{sl} for an
inhomogeneous gravitational system. This theory was further discussed
by Chavanis \cite{quasi} in an attempt to make a link with the
Maximum Entropy Production Principle.  We shall give here a simple
account of the quasilinear theory. Further details can be found in
Refs. \cite{kp,sl,quasi}. Our objective is to obtain an expression for
the diffusion current ${\bf J}$ by working directly on the
Vlasov-Poisson system, i.e. without {\it assuming} that the entropy
increases as is done in the thermodynamical approach.

Since the diffusion current ${\bf J}_{f}=\overline{\tilde f\tilde {\bf
F}}$ is related to the fine-grained fluctuations of the distribution
function, any systematic calculation starting from the Vlasov equation
(\ref{Vlasov}) must necessarily introduce an evolution equation for
$\tilde f$. This equation is simply obtained by substracting equation
(\ref{E15}) from equation (\ref{Vlasov}). This yields
\begin{equation}
\label{E29}
{\partial \tilde{f}\over\partial t}+{\bf v}{\partial \tilde{f}\over \partial {\bf r}}+\overline{\bf F}{\partial \tilde{f}\over\partial {\bf v}}=-\tilde{\bf F}{\partial \overline{f}\over\partial {\bf v}}-\tilde {\bf F}{\partial \tilde f\over\partial {\bf v}}+\overline{\tilde {\bf F}{\partial \tilde f\over\partial {\bf v}}}.
\end{equation}
The essence of the quasilinear theory is to assume that the
fluctuations are weak and neglect the nonlinear terms in equation (\ref{E29})
altogether. In that case, equations (\ref{E15}) and (\ref{E29}) reduce
to the coupled system
\begin{equation}
\label{E30}
{\partial \overline{f}\over\partial t}+L\overline{f}=-{\partial \over \partial {\bf v}}\overline{\tilde {\bf F}\tilde f},
\end{equation}
\begin{equation}
\label{E31}
{\partial \tilde{f}\over\partial t}+L\tilde{f}=-\tilde {\bf F}{\partial\overline{f}\over\partial {\bf v}},
\end{equation}
where $L={\bf v}{\partial\over\partial {\bf r}}+\overline{\bf F}{\partial\over\partial {\bf v}}$ is the advection operator in phase space.  Physically, these equations describe the coupling between a subdynamics (here the small scale fluctuations $\tilde f$) and a macrodynamics (described by the coarse-grained distribution function $\overline{f}$). 

Introducing the Greenian
\begin{equation}
\label{E32}
G(t_{2},t_{1})\equiv \exp \Biggl \lbrace -\int_{t_{1}}^{t_{2}}dt L(t)\Biggr \rbrace,
\end{equation}
we can immediately write down a formal solution of equation (\ref{E31}), namely
\begin{equation}
\label{E33}
\tilde f ({\bf r},{\bf v},t)=G(t,0)\tilde f ({\bf r},{\bf v},0)-\int_{0}^{t}ds G(t,t-s)\tilde {\bf F}({\bf r},t-s){\partial\overline{f}\over\partial {\bf v}}({\bf r},{\bf v},t-s).
\end{equation}
Although very compact, this formal expression is in fact extremely
complicated. Indeed, all the difficulty is encapsulated in the
Greenian $G(t,t-s)$ which supposes that we can solve the smoothed-out
Lagrangian flow
\begin{equation}
\label{E34}
{d{\bf r}\over dt}={\bf v},\qquad {d{\bf v}\over dt}=\overline{\bf F},
\end{equation} 
between $t$ and $t-s$. In practice, this is impossible and we will
have to make some approximations.

The objective now is to substitute the formal result (\ref{E33}) back
into equation (\ref{E30}) and make a closure approximation in order to
obtain a self-consistant equation for $\overline{f}({\bf r},{\bf
v},t)$.  If the fluctuating force $\tilde{\bf F}$ were external to the
system, we would simply obtain a diffusion equation
\begin{equation}
\label{E35}
{\partial \overline{f}\over\partial t}+L\overline{f}= {\partial\over\partial v^{\mu}}\biggl (D^{\mu\nu}{\partial\overline{f}\over\partial v^{\nu}}\biggr ),
\end{equation}
with a diffusion coefficient given by a Kubo formula
\begin{equation}
\label{E36}
D^{\mu\nu}=\int_{0}^{t} ds \overline{\tilde F^{\mu}({\bf r},t)\tilde F^{\nu}({\bf r},t-s)}.
\end{equation}
However, in the case of the Vlasov-Poisson system, the gravitational
force is in fact produced by the distribution of matter itself and
this coupling will give rise to a friction term in addition to the
pure diffusion. Indeed, we have
\begin{equation}
\label{E37}
\tilde {\bf F}({\bf r},t)=\int {\bf F}({\bf r}'\rightarrow {\bf r})\tilde f ({\bf r}',{\bf v'},t)d^{3}{\bf r}'d^{3}{\bf v}',
\end{equation}
where 
\begin{equation}
\label{E38}
{\bf F}({\bf r}'\rightarrow {\bf r})=G {{\bf r}'-{\bf r}\over |{\bf r}'-{\bf r}|^{3}},
\end{equation}
represents the force created by a (field) star in ${\bf r'}$ on a
(test) star in ${\bf r}$ (Newton's law). Therefore, considering
equations (\ref{E33}) and (\ref{E37}), we see that the fluctuations of
the distribution function $\tilde f ({\bf r},{\bf v},t)$ are given by
an iterative process: $\tilde f (t)$ depends on $\tilde {\bf F}(t-s)$
which itself depends on $\tilde f (t-s)$ etc... We shall solve this
problem perturbatively in an expansion in powers of the gravitational
constant $G$. This is the equivalent of the ``weak coupling
approximation'' in plasma physics.  To order $G^{2}$, we obtain after
some rearrangements
\begin{eqnarray}
\label{E39}
{\partial \overline{f}\over\partial t}+L\overline{f}={\partial \over \partial v^{\mu}}\int_{0}^{t} ds\int d^{3}{\bf r'}d^{3}{\bf v'}d^{3}{\bf r''}d^{3}{\bf v''} F^{\mu}({\bf r}'\rightarrow {\bf r})G'(t,t-s)G(t,t-s)\nonumber\\ 
\times\biggl\lbrace F^{\nu}({\bf r}''\rightarrow {\bf r})\overline{ \tilde f({\bf r}',{\bf v}',t-s)\tilde f({\bf r}'',{\bf v}'',t-s)} {\partial \overline{f}\over \partial v^{\nu}}({\bf r},{\bf v},t-s)\nonumber\\
+F^{\nu}({\bf r}''\rightarrow {\bf r}')\overline{\tilde f({\bf r},{\bf v},t-s)\tilde f({\bf r}'',{\bf v}'',t-s)} {\partial \overline{f}\over \partial v'^{\nu}}({\bf r}',{\bf v}',t-s) \biggr \rbrace.\nonumber\\
\end{eqnarray}
In this expression, the Greenian $G$ refers to the fluid particle
${\bf r}(t),{\bf v}(t)$ and the Greenian $G'$ to the fluid particle
${\bf r}'(t),{\bf v}'(t)$. To close the system, it remains for one to
evaluate the correlation function $\overline{\tilde f({\bf r},{\bf
v},t)\tilde f({\bf r'},{\bf v}',t)}$. We shall assume that the mixing
in phase space is sufficiently efficient so that the scale of the
kinematic correlations is small with respect to the coarse-graining
mesh size. In that case,
\begin{equation}
\label{E40}
\overline{\tilde f({\bf r},{\bf v},t)\tilde f({\bf
r'},{\bf v'},t)}=\epsilon_{r}^{3}\epsilon_{v}^{3} \delta({\bf r}-{\bf r}')\delta ({\bf v}-{\bf v}') \overline{\tilde f^{2}}({\bf r},{\bf v},t),
\end{equation}
where $\epsilon_{r}$ and $\epsilon_{v}$ are the resolution scales in
position and velocity respectively. Now,
\begin{equation}
\label{E41}
 \overline{\tilde f^{2}}=\overline{(f-\overline{f})^{2}}=\overline{f^{2}}-\overline{f}^{2}.
\end{equation}
We shall assume for simplicity that the initial condition in phase
space consists of a patch of uniform distribution function
($f=\eta_{0}$) surrounded by vaccum ($f=0$). This is the two-levels
approximation already considered in sections \ref{LB} and
\ref{MEPP}. In that case $\overline{f^{2}}=\overline{\eta_{0}\times
f}=
\eta_{0}\overline{f}$ and, therefore,
\begin{equation}
\label{E42}
\overline{\tilde f({\bf r},{\bf v},t)\tilde f({\bf
r'},{\bf v'},t)}=\epsilon_{r}^{3}\epsilon_{v}^{3} \delta({\bf r}-{\bf r}')\delta ({\bf v}-{\bf v}') \overline{f} (\eta_{0}-\overline{f}).
\end{equation}
Substituting this expression in equation (\ref{E39}) and carrying out
the integrations on ${\bf r}''$ and ${\bf v}''$, we obtain
\begin{eqnarray}
\label{E43}
{\partial \overline{f}\over\partial t}+L\overline{f}=\epsilon_{r}^{3}\epsilon_{v}^{3}{\partial \over \partial v^{\mu}}\int_{0}^{t} ds\int d^{3}{\bf r'}d^{3}{\bf v'} F^{\mu}({\bf r}'\rightarrow {\bf r})_{t} F^{\nu}({\bf r}'\rightarrow {\bf r})_{t-s}\nonumber\\ 
\times\biggl\lbrace  \overline{f}'(\eta_{0}-\overline{f}'){\partial \overline{f}\over \partial v^{\nu}}
- \overline{f}(\eta_{0}-\overline{f}) {\partial \overline{f}'\over \partial v'^{\nu}} \biggr \rbrace_{t-s}.
\end{eqnarray}
We have written $\overline{f}'_{t-s}\equiv \overline{f}({\bf
r}'(t-s),{\bf v}'(t-s),t-s)$, $\overline{f}_{t-s}\equiv
\overline{f}({\bf r}(t-s),{\bf v}(t-s),t-s)$, $F^{\mu}({\bf
r}'\rightarrow {\bf r})_{t}\equiv F^{\mu}({\bf r}'(t)\rightarrow {\bf
r}(t))$ and $F^{\nu}({\bf r}'\rightarrow {\bf r})_{t-s}\equiv
F^{\nu}({\bf r}'(t-s)\rightarrow {\bf r}(t-s))$ where ${\bf r}(t-s)$
and ${\bf v}(t-s)$ are the position and velocity at time $t-s$ of the
stellar fluid particle located in ${\bf r}={\bf r}(t)$, ${\bf v}={\bf v}(t)$
at time $t$. It is determined by the characteristics (\ref{E34}) of
the smoothed-out Lagrangian flow.

Equation (\ref{E43}) is a non Markovian integrodifferential equation:
the value of $\overline{f}$ in ${\bf r}$, ${\bf v}$ at time $t$
depends on the value of the { whole} field $\overline{f}({\bf r}',{\bf
v}',t-s)$ at { earlier times}. If the decorrelation time $\tau$ is
short, we can make a {\it Markov approximation} and replace the
bracket at time $t-s$ by its value taken at time $t$. Noting
furthermore that the integral is dominated by the contribution of
field stars close to the star under consideration (i.e. when ${\bf
r}'\rightarrow {\bf r}$), we shall make a {\it local approximation}
and replace $ \overline{f}'(\eta_{0}-\overline{f}')$ and $ {\partial
\overline{f}'\over \partial v'^{\nu}}$ by their values taken at ${\bf
r}$. In that case, the foregoing equation simplifies in
\begin{eqnarray}
\label{E44}
{\partial \overline{f}\over\partial t}+L\overline{f}=\epsilon_{r}^{3}\epsilon_{v}^{3}{\partial \over \partial v^{\mu}}\int_{0}^{t} ds\int d^{3}{\bf r'}d^{3}{\bf v'} F^{\mu}({\bf r}'\rightarrow {\bf r})_{t} F^{\nu}({\bf r}'\rightarrow {\bf r})_{t-s}\nonumber\\ 
\times\biggl\lbrace  \overline{f}'(\eta_{0}-\overline{f}'){\partial \overline{f}\over \partial v^{\nu}}
- \overline{f}(\eta_{0}-\overline{f}) {\partial \overline{f}'\over \partial v'^{\nu}} \biggr \rbrace_{t},
\end{eqnarray}
where, now, $\overline{f}'=\overline{f}({\bf r},{\bf v}',t)$. The
explicit reference to the past evolution of the system is only
retained in the memory function 
$$\int_{0}^{t} ds \int d^{2}{\bf r}'
F^{\mu}({\bf r}'\rightarrow {\bf r})_{t} F^{\nu}({\bf r}'\rightarrow
{\bf r})_{t-s}.$$ 
This function can be calculated explicitly if we
assume that, between $t-s$ and $t$, the stars follow linear
trajectories, so that ${\bf v}(t-s)={\bf v}$ and ${\bf r}(t-s)={\bf
r}-{\bf v}s$ \cite{balescu}. This leads to the generalized Landau
equation
\begin{eqnarray}
\label{E45}
{\partial \overline{f}\over\partial t}+L\overline{f}={\partial\over\partial v^{\mu}}\int d^{3}{\bf v}' K^{\mu\nu}\biggl\lbrace \overline{f}'(\eta_{0}-\overline{f}'){\partial\overline{f}\over\partial v^{\nu}}-\overline{f}(\eta_{0}-\overline{f}){\partial\overline{f}'\over\partial v'^{\nu}}\biggr\rbrace,
\end{eqnarray}
where $K^{\mu\nu}$ is the tensor
\begin{eqnarray}
\label{E46}
K^{\mu\nu}=2\pi G^{2}\epsilon_{r}^{3}\epsilon_{v}^{3}\ln\Lambda {1\over u}\biggl (\delta^{\mu\nu}-{u^{\mu}u^{\nu}\over u^{2}}\biggr ),
\end{eqnarray}
and ${\bf u}={\bf v}'-{\bf v}$, $\ln\Lambda=\ln
(R/\epsilon_{r})$. This equation applies to inhomogeneous systems but,
as a result of the local approximation, the effect of inhomogeneity is
only retained in the advective term. Equation (\ref{E45}) is very
similar to the well-known Landau equation of collisional
self-gravitating systems and electric charges \cite{Kandrup}. There
are nevertheless two important differences: (i) The friction term is
non linear in $\overline{f}$, accounting for the degeneracy discovered
by Lynden-Bell at equilibrium. (ii) The diffusion coefficient is
proportional to the mass $\eta_{0}\epsilon_{r}^{3}\epsilon_{v}^{3}$ of
a macrocell completely filled by the phase fluid instead of the mass
$m$ of an individual star in the ordinary Landau equation. In general
$\eta_{0}\epsilon_{r}^{3}\epsilon_{v}^{3}\gg m$ so that the relaxation
by phase mixing is much more rapid than the collisional
relaxation. From the above theory, we find that the collisionless
relaxation operates on a time scale $\sim t_{D}$, the dynamical time,
whereas the collisional relaxation operates on a time scale
$t_{coll}\sim {N\over\ln N}t_{D}\gg t_{D}$ where $N\sim 10^{12}$ is the number
of stars in the galaxy. Therefore, the relaxation by phase mixing
really corresponds to a ``violent relaxation'' \cite{lb}.

It can be shown that the generalized Landau equation (\ref{E45})
conserves mass and energy \cite{balescu}. In fact, as a result of the
local approximation, the energy is conserved locally, as if the system
were homogeneous, and we have
\begin{eqnarray}
\label{E47}
\int \biggl ({\partial\overline{f}\over\partial t}\biggr )_{c.g.}{v^{2}\over 2}d^{3}{\bf v}=\int {\bf J}_{f}{\bf v}d^{3}{\bf v}=0, \qquad  \forall {\bf r}.
\end{eqnarray}
It is also easy to show that equation (\ref{E45}) satisfies a
H-theorem ($\dot S\ge 0$) for the Fermi-Dirac entropy
(\ref{E9}). When a stationary state is reached $\dot S=0$ and the
Fermi-Dirac distribution (\ref{E10}) is obtained, in agreement with
Lynden-Bell's statistical theory \cite{lb}. This provides therefore
another way of justifying his results from a dynamical point of view
which does not explicitly rely on a maximization of entropy (the
$H$-theorem is not assumed but {\it derived} from the kinetic
theory). It can be noted that these properties result from the {\it
symmetry} of the Landau collision term and not from Lagrange
multipliers like in the thermodynamical approach. This is more
satisfactory on a physical point of view.

It is important to stress, however, that this quasilinear theory
cannot describe the early, very non linear and very chaotic, stages of
the ``violent relaxation''. Indeed, the detailed study of Severne \&
Luwel \cite{sl} reveals that the various approximations introduced in
the theory make equation (\ref{E45}) applicable only for $t\gg t_{D}$,
where $t_{D}$ is the dynamical time. Since the relaxation time is
precisely of order $t_{D}$, the quasilinear theory is only {\it
marginally} applicable and can describe, at most, the late quiescent
phases of the relaxation, when the fluctuations have weaken. By
contrast, it is plausible that equation (\ref{E25}) is
more general and more appropriate to the context of ``violent
relaxation'' since it results from a thermodynamical approach which
exploits at best the chaoticity of the system and the complete lack of
information that we have to face at small scales. As a clear
difference, it should be noted that the MEPP takes into account only
the {\it global} conservation of energy (the Lagrange multiplier
$\beta (t)$ is a kind of inverse average temperature determined by an
integration over the whole system) whereas the approximations
introduced in the kinetic model lead to a {\it local} conservation of
energy. This strong locality cannot account for the rather collective
processes which are involved in the violent relaxation and may unveil
a failure of the quasilinear theory. However, in section \ref{TM}, we
show that the two approaches are consistant if we are close to
equilibrium and we use the quasilinear theory as a model to determine
an explicit expression for the diffusion coefficient that appears in
equation (\ref{E25}).

\section{Truncated models for collisionless stellar systems}
\label{TM}

The equations provided by the MEPP and by the quasilinear theory have
a very different mathematical structure. The relaxation equation
(\ref{E25}) is a partial differential equation whereas the generalized
Landau equation (\ref{E45}) is an integrodifferential equation: the
value of $\overline{f}({\bf v})$ at time $t+dt$ depends explicitly on
the {\it whole} distribution of velocities $\overline{f}({\bf v}')$ at
time $t$ through an integration over ${\bf v}'$. The usual way to
transform an integrodifferential equation into a partial differential
equation is to make a guess for the function appearing in the integral
and refine the initial guess by successive iterations. In practice, we
simply make {\it one} sensible guess. Therefore, if we are close to
equilibrium (and this is in fact dictated by the conditions of
validity of the quasilinear theory), it seems natural to replace the
distribution function $\overline{f}'$ by the Fermi-Dirac distribution
\begin{equation}
\label{E48}
\overline{f}'={\eta_{0}\over 1+\lambda e^{\beta\eta_{0}{v'^{2}\over 2}}}.
\end{equation}
In more physical terms, this amounts to a ``thermal bath
approximation'': the stars have not yet relaxed completely, but when
we focus on the relaxation of a given stellar fluid particle
(described by $\overline{f}$) we can consider, in a first
approximation, that the rest of the system (described by
$\overline{f}'$) is at equilibrium. With this thermal bath
approximation, the generalized Landau equation (\ref{E45}) reduces to
the nonlinear partial differential equation \cite{quasi}:
\begin{equation}
\label{E49}
{\partial \overline{f}\over \partial t}+{\bf  v}{\partial\overline{f}\over\partial{\bf  r}}+\overline{\bf F}{\partial\overline{f}\over\partial{\bf  v}}={\partial\over\partial  {\bf v}}\biggl\lbrack D \biggl ({\partial\overline{f}\over\partial   {\bf v}}+\beta\overline{f} (\eta_{0}-\overline{f})  {\bf v}\biggr )\biggr\rbrack,
\end{equation}
with a diffusion coefficient
\begin{equation}
\label{E51}
D={16\sqrt 2\pi^{2}G^{2}\epsilon_{r}^{3}\epsilon_{v}^{3}\ln\Lambda\over\eta_{0}^{1/2}\beta^{5/2} v^{3}}\int_{0}^{\beta\eta_{0}{v^{2}\over 2}}{\sqrt x\over 1+\lambda e^{x}}dx.
\end{equation}
Equation (\ref{E49}) is precisely the equation derived from the
MEPP. Together with the explicit expression of the diffusion
coefficient (\ref{E51}) it provides a self-consistent model for the
``coarse-grained'' dynamics of collisionless stellar systems
experiencing violent relaxation. More general equations including
anisotropic effects can also be obtained from this formalism
\cite{quasi}.

We shall now assume that the galaxy is not isolated but subject to the
tides of other systems. In that case, high energy stars that have
elongated orbits are removed by the gravity of these objects. We seek
therefore a stationary solution of equation (\ref{E49}) satisfying the
boundary condition $\overline{f}(\epsilon_{m})=0$, where $\epsilon_{m}$
is the escape energy above which $\overline{f}=0$. This solution will
provide a truncated distribution function with a finite mass
\cite{quasi}. In fact, this problem was already tackled by Lynden-Bell
in his seminal paper \cite{lb}. He proposed to describe the evolution
of the coarse-grained distribution function $\overline{f}$ by the
ordinary Fokker-Planck equation (\ref{E26}) with the heuristic
argument that the fluctuations of the gravitational potential during
violent relaxation play the same role as collisions. With the
additional (heuristic) argument that $D\sim {1\over v^3}$ for large
velocities, he could obtain a stationary solution of a Michie-King
type \cite{bt}:
\begin{eqnarray}
\overline{f}=\biggl\lbrace \begin{array}{cc}
A (e^{-\beta\eta_{0} \epsilon}-e^{-\beta\eta_{0} \epsilon_{m}}) & \epsilon\le \epsilon_{m}, \\
0 & \epsilon\ge \epsilon_{m}.
\end{array}
\label{E52}
\end{eqnarray} 
Since stars with $\epsilon\ge \epsilon_{m}$ are removed by the tidal
field, this distribution function provides a depletion of high energy
states and solves the infinite mass problem. 

Our present approach
justifies the two phenomenological arguments of Lynden-Bell since
equation (\ref{E49}) reduces to the Fokker-Planck equation if we
assume $\overline{f}\ll \eta_{0}$ (no degeneracy) and equation
(\ref{E51}) gives a diffusion coefficient $D\sim {1\over v^3}$ for
large velocities. There is however a kind of ``gap'' in Lynden-Bell's
approach since equation (\ref{E52}) does not reduce to the Fermi-Dirac
statistics (\ref{E10}) in the limit of low energies. Using the more
general equation (\ref{E49}) which properly accounts for degeneracy
effects, we now try to build up a ``one piece'' distribution function
which makes the bridge between Lynden-Bell's statistics (\ref{E10})
for $\epsilon\ll \epsilon_{m}$ and the Michie-King model (\ref{E52})
for $\epsilon\sim\epsilon_{m}$.

During the stage of violent relaxation, the stars extract their energy
from the rapid fluctuations of the gravitational field. By this
process, some stars may acquire very high energies and escape from the
system (being ultimately captured by the gravity of other
systems). For these stars, $D={K\over v^3}$ is a good approximation
and equation (\ref{E49}) reads
\begin{equation}
\label{E53}
{\partial \overline{f}\over \partial t}+{\bf  v}{\partial\overline{f}\over\partial{\bf  r}}+\overline{\bf F}{\partial\overline{f}\over\partial{\bf  v}}={\partial\over\partial {\bf v}}\Biggl\lbrack {K\over v^3}\Biggl ( {\partial\overline{f}\over \partial {\bf v}}+\beta\overline{f}(\eta_{0}-\overline{f}){\bf v}\Biggr )\Biggr \rbrack.
\end{equation}
We seek a stationary solution of the form $\overline{f}=\overline{f}(\epsilon)$. Using the identity ${\partial\over\partial {\bf v}}({{\bf v}\over v^3})=0$ (valid for sufficiently large $|{\bf v}|$), we obtain
\begin{equation}
\label{E54}
0={d\over d \epsilon}\Biggl\lbrack K\Biggl ( {d\overline{f}\over d\epsilon}+\beta\overline{f}(\eta_{0}-\overline{f})\Biggr )\Biggr \rbrack,
\end{equation}
or, equivalently,
\begin{equation}
\label{E55}
{d\overline{f}\over d \epsilon}+\beta\overline{f}(\eta_{0}-\overline{f})=-J,
\end{equation}
where $J$ is a constant of integration. If $J=0$, we recover
Lynden-Bell's distribution function (\ref{E10}). If $J\neq 0$,
equation (\ref{E55}) accounts physically for an escape of stars at a
constant rate $J$. The system is therefore not truly static since it
looses gradually stars but we may consider that the galaxy passes by a
succession of quasi-stationary states which are solution of equation
(\ref{E55}). As stated previously, this equation is valid only for
high energy stars. For lower energies, the system has settled down in
a pure equilibrium state and $J=0$, leading to Lynden-Bell's
distribution.

Our goal, now, is to solve equation (\ref{E55}). Put under the form
\begin{equation}
\label{E56}
{d\overline{f}\over d\epsilon}+\beta\eta_{0}\overline{f}-\beta\overline{f}^{2}+J=0,
\end{equation}
we recognize a Riccatti equation \cite{Ince}. With the change of
variables $\overline{f}=-{1\over\beta}{u'\over u}$, it can be
converted into a linear partial differential equation of second order
\begin{equation}
\label{E57}
{d^2u\over d\epsilon^2}+\eta_{0}\beta {du\over d\epsilon}-J\beta u=0.
\end{equation}
The associated characteristic polynomial $x^2+\eta_{0}\beta x-J\beta=0$ has a strictly positive discriminant $\Delta=\eta_{0}^2\beta^2+4J\beta>0$. Therefore, the general solution of equation (\ref{E57}) is
\begin{equation}
\label{E58}
u=e^{-{\eta_{0}\beta\over 2}\epsilon}( A e^{{\delta\over 2}\epsilon}+B e^{-{\delta\over 2}\epsilon}),
\end{equation}
where $\delta=\sqrt{\Delta}$ and $A$, $B$ are constants of integration. The solution of equation (\ref{E56}) is therefore
\begin{equation}
\overline{f}={1\over 2\beta}{A(\eta_{0}\beta-\delta)e^{{\delta\over 2}\epsilon}+B (\eta_{0}\beta+\delta)e^{-{\delta\over 2}\epsilon}\over A e^{{\delta\over 2}\epsilon}+B e^{-{\delta\over 2}\epsilon}}.
\label{E59}
\end{equation}
Setting $\lambda=A/B$ ($B\neq 0$ otherwise $\overline{f}$ would be constant), it can be rewritten
\begin{equation}
\overline{f}={1\over 2\beta}{\lambda(\eta_{0}\beta-\delta)+(\eta_{0}\beta+\delta)e^{-\delta\epsilon}\over \lambda+e^{-\delta\epsilon}}.
\label{E60}
\end{equation}
This distribution function vanishes at the escape energy $\epsilon=\epsilon_{m}$ defined by
\begin{equation}
\lambda (\eta_{0}\beta-\delta)+(\eta_{0}\beta+\delta)e^{-\delta\epsilon_{m}}=0.
\label{E61}
\end{equation}
With this new variable, we obtain the result
\begin{equation}
\overline{f}=\lambda\eta_{0} {e^{-\delta\epsilon}-e^{-\delta\epsilon_{m}}\over (\lambda-e^{-\delta\epsilon_{m}})(\lambda+e^{-\delta\epsilon})},
\label{E62}
\end{equation}
where $\delta$ is a solution of
\begin{equation}
\delta=\eta_{0}\beta {\lambda+e^{-\delta\epsilon_{m}}\over \lambda-e^{-\delta\epsilon_{m}}}.
\label{E63}
\end{equation}
Now, for the cases of physical interest $\lambda\gg
e^{-\beta\eta_{0}\epsilon_{m}}$, which means that degeneracy is
negligible for energies close to the escape energy $\epsilon_{m}$. In
that case, $\delta\simeq \eta_{0}\beta$ and we obtain the final result
\cite{quasi}:
\begin{equation}
\overline{f}=\eta_{0}{e^{-\beta\eta_{0}\epsilon}-e^{-\beta\eta_{0}\epsilon_{m}}\over \lambda+e^{-\beta\eta_{0}\epsilon}}.
\label{E64}
\end{equation}

The previous calculation is justified as long as $D\neq 0$,
corresponding to relatively strong fluctuations. When the fluctuations
die down at the end of the relaxation, the diffusivity $D$ and
therefore the diffusion current $|{\bf J}|$ go to zero. There is no
more evaporation but the distribution function (\ref{E64}) is
maintained as a stationary solution of the Vlasov equation. When
$\epsilon\sim\epsilon_{m}$, we recover the Michie-King model
(\ref{E52}) and when $\epsilon\ll\epsilon_{m}$ equation (\ref{E64})
reduces to the Fermi-Dirac distribution function
(\ref{E10}). Therefore, the distribution function (\ref{E64}) connects
{\it continuously} the two limits considered by Lynden-Bell \cite{lb}
and can serve as a relevant model for (possibly degenerate)
collisionless stellar systems. In particular, this distribution
function could describe galactic halos limited in extension as a
consequence of tidal interactions with other systems
\cite{ruffini}. It could also be used to calculate realistic
equilibrium states of collisionless stellar systems without the
artifice of a material box. However, the main results of the box model
\cite{cs} should not be dramatically altered.

The previous model describes self-gravitating systems subject to tidal
forces. This form of confinement was first introduced in the case of
globular clusters tidally trucated by a nearby galaxy \cite{bt}. This
model can also describe a fraction of elliptical galaxies living in a
rich environement. However, for the majority of elliptical galaxies,
tidal forces are weak and the galaxy can be considered as
isolated. Now, for isolated systems, other processes can account for
``incomplete relaxation'' and lead to different truncated models. Starting
from equation (\ref{E36}), we can show that the diffusion coefficient
is proportional to the fluctuations of the distribution function
integrated over the velocity:
\begin{eqnarray}
D^{\mu\nu}=2\pi G^{2}\epsilon_{r}^{3}\epsilon_{v}^{3}\ln\Lambda\int  {1\over u}\biggl (\delta^{\mu\nu}-{u^{\mu}u^{\nu}\over u^{2}}\biggr )(\overline{f'^{2}}-\overline{f}'^{2}) d^{3}{\bf v}'.
\label{E65}
\end{eqnarray}
Since these fluctuations rapidly decay as we depart from the
relatively well-mixed central region of the galaxy, the diffusion
current decreases also and this results in a confinement of the
distribution function. It is expected therefore that the relaxation
will be effective only in a limited region of space where
$\overline{f^{2}}-\overline{f}^{2}$ is sufficiently large. On the
other hand, as the system develops finer and finer filaments during
the mixing process, the diffusion coefficient is expected to decrease
in time. The diffusion current takes therefore increasingly small
values and the relaxation is slowed down. This decay may be quite
rapid and a quantitative treatment of this effect would require a
better understanding of the correlation function $\overline{ \tilde f
({\bf r},{\bf v},t)\tilde f ({\bf r'},{\bf v'},t)}$ whose expression
was simply postulated in section \ref{QT}. The development of these
ideas will lead to other truncated models probably closer to those
introduced phenomenologically by, e.g., Stiavelli \& Bertin \cite{sb},
Tremaine \cite{t} and Hjorth \& Madsen \cite{hjorth}. In particular,
the above discussion is quite consistent with the scenario of
incomplete violent relaxation developed by Hjorth \& Madsen. These
authors introduce a two-step process: (i) in a first step, they assume
that violent relaxation proceeds to completion in a {\it finite}
spatial region, of radius $r_{max}$, which represents roughly the core
of the galaxy (where the fluctuations are important). At this stage,
the escape energy represents no special threshold so that negative as
well as positive energy states are populated in that region. (ii) After
the relaxation process is over, positive energy particles leave the
system and particles with $\Phi(r_{max})<\epsilon<0$ move in orbits
beyond $r_{max}$, thereby changing the distribution function to
something significantly `thinner' than a Boltzmann distribution. The
crucial point to realize is that the differential energy distribution
$N(\epsilon)$, where $N(\epsilon)d\epsilon$ is the number of stars
with energy between $\epsilon$ and $\epsilon+d\epsilon$, will be
discontinuous at the escape energy $\epsilon=0$ since there is a
finite number of particles within $r_{max}$ after the relaxation
process. It can be shown that this discontinuity implies necessarily
that $f(\epsilon)\sim (-\epsilon)^{5/2}$ for $\epsilon\rightarrow
0^{-}$ \cite{jaffe}. Therefore, the truncated model consistent with
this scenario is \cite{hjorth}:
\begin{eqnarray}
\overline{f}=\Biggl\lbrace \begin{array}{cc}
A e^{-\beta\eta_{0} \epsilon} & \epsilon< \epsilon'; \\
B \ (-\epsilon)^{5/2} & \epsilon'\le\epsilon<0; \\
0  & \epsilon\ge 0
\end{array}
\label{hm}
\end{eqnarray}  
This truncated model corresponds formally to a composite
configuration made of an isothermal core and a polytropic envelope
with index $n=4$. Of course, if the core is degenerate, the Boltzmann
distribution for $\epsilon< \epsilon'$ must be replaced by the
Fermi-Dirac one. This truncated model gives very good fit with
elliptical galaxies and can reproduce the $R^{1/4}$ law \cite{hjorth}.
The box model of section \ref{FD} can be considered as a simple
approximation of this more realistic model, the ``box'' playing the
role of the envelope. In fact, the scenario developed by Hjorth \&
Madsen is almost equivalent to complete violent relaxation in a finite
container with the container removed after the relaxation. This
scenario solves the infinite mass problem and rehabilitates the use of statistical mechanics to understand the structure of
elliptical galaxies.

\section{Conclusion}
\label{conclusion}

We have described in this paper the process of violent relaxation in
stellar systems from the viewpoint of statistical mechanics, as
originally introduced by Lynden-Bell \cite{lb}. Lynden-Bell proposed
that the structure of galaxies could result from a law of chaos: there
is a total lack of information at small scales since the stars have
complicated orbits, but the exciting phenomenon is that {\it
microscopic} disorder leads to {\it macroscopic} order. The same ideas
of statistical mechanics have been introduced in two-dimensional
turbulence described by the Euler equation \cite{rs1,miller,csr,csfluide}
to explain the formation and maintenance of large scale coherent
vortices like the Great Red Spot of Jupiter or the cyclones and
anticyclones that populate the earth atmosphere
\cite{shallow,bouchet1,bouchet2}.  The formation of
self-organized vortices in two-dimensional turbulence can also have
applications in the context of planet formation where large-scale
vortices present in the solar nebula could efficiently trap dust
particles to form the planetesimals and the planets (see a complete
list of references in Chavanis \cite{cplanetes}). In fact, the
statistical mechanics of the 2D Euler equation is equivalent to the
theory of Lynden-Bell applying to the Vlasov equation. In a sense, the
Vlasov equation is just the Euler equation for a ``fluid'' evolving in
a six-dimensional phase space.  In this analogy
\cite{cthese,cfloride,japon}, the vorticity and the stream function in
2D turbulence play the same role as the distribution function and the
gravitational field in galaxies. This analogy concerns not only the
equilibrium states (the formation of large-scale structures) but also
the relaxation towards equilibrium \cite{csr} and the statistics of
fluctuations \cite{csire}. A kinetic theory of two-dimensional
vortices can be developed in analogy with stellar dynamics
\cite{kin2}. In this kinetic theory, a point vortex experiences a
diffusion process and a ``systematic drift''. This ``systematic
drift'' \cite{drift} is the counterpart of the ``dynamical friction''
\cite{chandrabrownien} experienced by a star as a result of close encounters.  
Relaxation equations analogous to equations (\ref{E25}) and
(\ref{E45}) have been derived in the context of vortex dynamics
\cite{rs,drift,kin1,kin2}. In addition, the problem of ``incomplete
relaxation'' is also encountered in 2D turbulence to explain the
confinement of a vortex (e.g., a dipole or a tripole) that forms after
a rapid merging \cite{csr,csfluide,rr}. It has been demonstrated
explicitly in two-dimensional turbulence (where the numerical
simulations are easier to implement) that the relaxation equations
derived from the MEPP and including a space dependant diffusion
coefficient related to the fluctuations of the vorticity (analogous to
Eq. (\ref{E65})) can account for this kinetic confinement \cite{rr}. We
believe that the relaxation equations proposed in the stellar context
should work as well. The statistical mechanics of continuous vorticity
fields also predicts a Fermi-Dirac distribution at equilibrium with
the same interpretation of the degeneracy as in Lynden-Bell's
theory. Although this degeneracy is hard to evidence for galaxies (and
remains controversal), it has been vindicated by various numerical
simulations and laboratory experiments of two-dimensional fluids. This
suggests that the degenerate version of the theory should also be used
in the stellar context.  If all these effects are taken into account
correctly, it is plausible that the statistical mechanics of 2D
vortices and self-gravitating systems has a chance to account for the
fascinating process of self-organization in nature.

\clearpage
\addcontentsline{toc}{section}{Index}
\flushbottom
\printindex

\end{document}